\documentclass[12pt]{iopart}
\usepackage{graphicx}%

\begin{document}
\title{Lack of static lattice distortion in Tb$_{2}$Ti$_{2}$O$_{7}$}
\author{Oren Ofer$^1$, Amit Keren$^1$, Chris Baines$^2$ and Jason S. Gardner$^3$}

\address{$^1$Physics Department, Technion, Israel Institute of
Technology, Haifa 32000, Israel}
\address{$^2$Laboratory for Muon-Spin Spectroscopy, Paul Scherrer
Institute, CH 5232 Villigen PSI, Switzerland}
\address{$^3$Indiana University, 2401 Milo B. Sampson Lane, Bloomington, IN
47408, USA}

\ead{keren@physics.technion.ac.il}
\begin{abstract}
We investigated the possibility of temperature dependent lattice
distortions in the pyrochlore compound Tb$_{2}$Ti$_{2}$O$_{7}$ by
measuring the internal magnetic field distribution, using muon
spin resonance, and comparing it to the susceptibility. The
measurements are done at temperatures as low as 70~mK and external
fields up to 6~kG. We find that the evolution of the width of the
field distribution can be explained by spin susceptibility only,
thus ruling out a temperature dependent hyperfine coupling. We
conclude that lattice deformations are absent in
Tb$_{2}$Ti$_{2}$O$_{7}$.
\end{abstract}
 \maketitle

Despite its short-range AFM correlations at temperatures lower
than 100K, the Tb$_{2}$Ti$_{2}$O$_{7}$ remains in a fluctuating
paramagnetic spin-liquid state down to 70mK \cite{GardnerPRL99}.
This very unusual state of matter has attracted considerable
attention of experimentalist and theorist alike. Recently it was
demonstrated that, under pressure \cite{MirebeauNature02} high
magnetic field \cite{Rule0601715} or both \cite{MirebeauPRL04}, Tb$_{2}$%
Ti$_{2}$O$_{7}$ does order magnetically. The possibility that this
order stems from magneto-elastic coupling was considered in all
cases \cite{MirebeauNature02,Rule0601715,MirebeauPRL04}. This type
of coupling allows the lattice to distort in order to relieve the
magnetic frustration, concomitantly lowering the total system
energy. Thus the magneto-elastic coupling, which is a small
perturbation to the spin Hamiltonian, can select one ground state
out of the macroscopically\ degenerate ground states
\cite{Theory,KerenPRL01}. In the most spectacular case the
magneto-elastic coupling leads to long range spin order
accompanied by a new lattice structure such as in
ZnCr$_{2}$O$_{4}$ \cite{LeePRL00} and CdCr$_2$O$_4$\cite{chung}.
An alternative and less dramatic possibility is a selection of a
ground state with short range spin order and short range lattice
deformation. In the latter case the original lattice structure is
preserved on the average. This might be the situation in
Y$_{2}$Mo$_{2}$O$_{7}$ where evidence for lattice deformations
were found by several methods \cite{KerenPRL01,Booth,eva}.

In this work we study the possible existence of magneto-elastic
coupling in Tb$_{2}$Ti$_{2}$O$_{7}$ using the muon spin resonance
($\mu$SR) technique. The basic idea is to investigate the nature
of the changes in the local environment of the muon as the
temperature decreases. We determine whether only the spin
polarization is changing or whether the lattice is involved as
well. Electronic spin polarization contributes to the shift of the
muon spin rotation frequency. Lattice distortions are responsible
for muon spin polarization relaxation. Comparing these two
quantities provides information on the presence or absence of
magneto-elastic coupling.

\begin{figure}
[ptb]
\begin{center}
\includegraphics{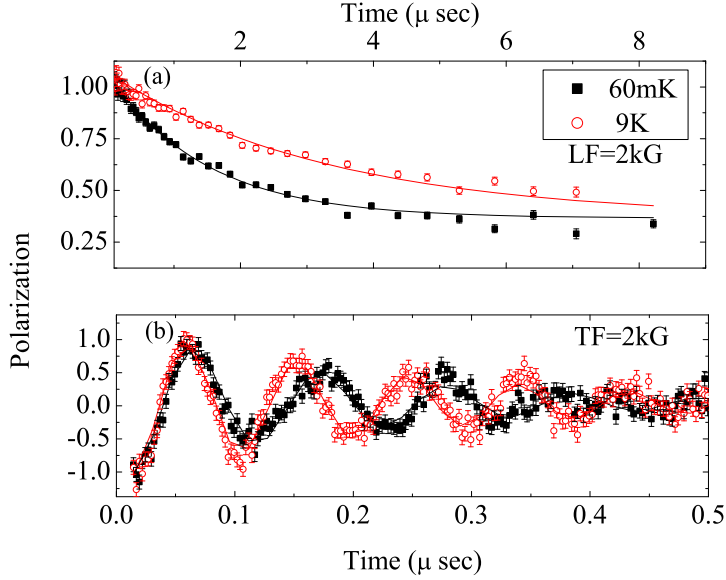}%
\caption{The time dependence of the muon polarization in the (a) longitudinal
and (b) transverse directions at an applied field of 2~kG and two different
temperatures. The time scale is different between the two directions. The
solid lines are fits to Eqs.~\ref{LFFit} and \ref{TFFit}.}%
\label{musraw}%
\end{center}
\end{figure}

Transverse [TF] and longitudinal field [LF] $\mu$SR measurements
were performed with powder samples on the GPS and LTF
spectrometers at Paul Scherrer Institute, Switzerland. The
measurements were carried out with the muon spin tilted by
$50^{0}$ relative to the direction of the applied magnetic fields,
and positron data were accumulated in both the forward-backward
(longitudinal) and the up-down (transverse) directions
simultaneously. This allowed us to determine both transverse and
longitudinal muon spin relaxations rates. In Fig. \ref{musraw} we
show the LF [panel (a)] and the TF [panel (b)] data at two
temperatures and applied field of 2kG. The TF data are shown in a
reference frame rotating at a field of 1.5kG. Several aspects can
be seen in the raw data: From the time scale it is clear that the
transverse relaxation is by far greater than the longitudinal one.
The longitudinal relaxation increases as the temperature
decreases, as was observed previously \cite{KerenPRL04}. Finally,
the transverse relaxation increases and the muon
rotation frequency decreases upon cooling.%

The $\mu$SR LF polarization is best described by the root exponential
\begin{equation}
P_{LF}(t)=A_{LF}\exp(-(t/T_{1})^{\frac{1}{2}})+B_{LF}\label{LFFit}%
\end{equation}
where the parameter $A_{LF}$ is set by taking into account the
tilt of the muon spin relative to the longitudinal magnetic field,
$T_{1}$ is the longitudinal relaxation time, $B_{LF}$ is the
background, and $t$ is time. Similarly, the TF polarization is
best fitted by a root exponential superimposed on a cosine
oscillation
\begin{equation}
P_{TF}(t)=A_{TF}\exp(-(t/T_{2})^{\frac{1}{2}})cos(\omega t+\phi)+B_{TF}%
.\label{TFFit}%
\end{equation}
Here $T_{2}$ is the transverse relaxation time. The other parameters have the
same mining as in Eq.~\ref{LFFit}. The quality of the fits is presented by the
solid lines in Fig.~\ref{musraw}.

\begin{figure}
[ptb]
\begin{center}
\includegraphics{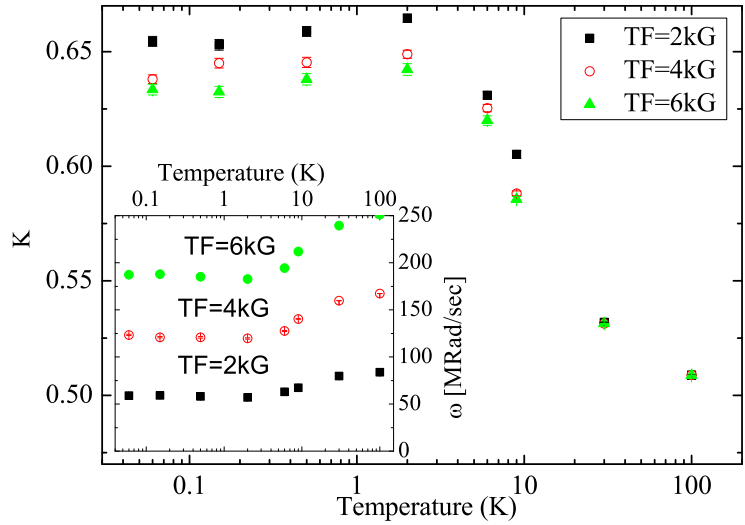}%
\caption{Inset: the muon rotation frequency as a function of temperature for
different applied fields. Main figure: the shift in the rotation frequency
defined in Eq.~\ref{Shift}.}%
\label{shift}%
\end{center}
\end{figure}

\begin{figure}
[ptb]
\begin{center}
\includegraphics{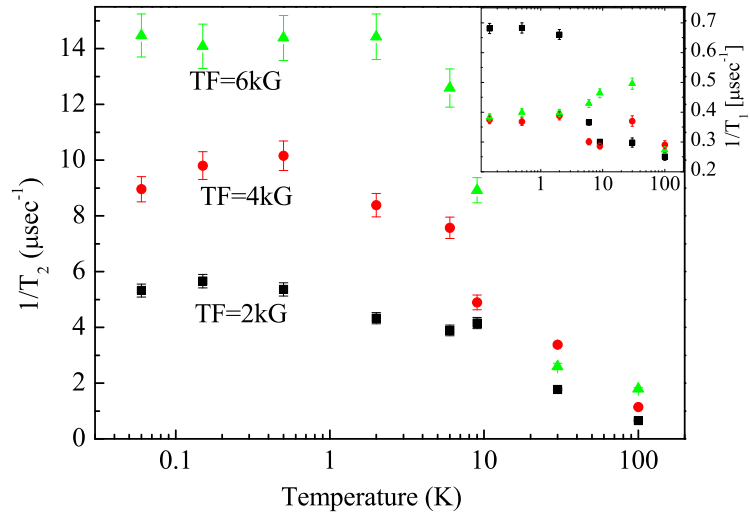}%
\caption{The muon spin transverse relaxation rate $T_{2}^{-1}$ as
a function of temperature for three different fields. Inset, the
muon spin longitudinal relaxation rates $T_{1}^{-1}$ at the same
temperatures and
fields.}%
\label{rlx2}%
\end{center}
\end{figure}

Data were collected in the temperature range 60~mK to 100~K and
three fields of 2, 4, and 6~kG. The frequency $\omega$ as a
function of temperature for the three different fields is depicted
in the inset of Fig.~\ref{shift}. The frequency shift,
\begin{equation}
K=(\omega_0-\omega)/\omega_{0}\label{Shift}%
\end{equation}
is shown in the main panel for the same temperatures and fields.
We define $\omega_0$ as the frequency of the free muon in the
rotating reference frame.  Similarly, we present $T_{2}^{-1}$ in
Fig. \ref{rlx2}, and in the inset $T_{1}^{-1}$. The most important
aspect of the data is that
all quantities saturate as the temperature decreases below $\sim$2K.%

The shift and LF relaxation have a simple interpretation. $K$ is a result of
the sample magnetization, and $T_{1}^{-1}$ stems from dynamic field
fluctuations. The relaxation $T_{2}^{-1}$ is a bit more involved. It is a
result of both static field inhomogeneities on the time scale of one muon spin turn,
and dynamically fluctuating fields.
However, since $T_{1}^{-1}$ is an order of magnitude lower than
$T_{2}^{-1}$, the contribution to the TF relaxation from dynamic fluctuations
is negligible. Therefore, $T_{2}^{-1}$ could be analyzed in terms of fields
inhomogeneities only.%

The $T_{2}^{-1}$ is related to lattice deformation via the muon coupling to
its neighboring spins and the system's susceptibility. To demonstrate this
relation let's assume for simplicity that the muon is coupled only to one
electronic spin $\mathbf{S}$ (extension to multiple couplings is trivial). In
this case the magnetic field experienced by the muon is a sum of the external
field $\mathbf{H}$ and the field from the neighboring electron $\mathbf{H}%
_{int}=g\mu_{B}\mathbf{AS}$ where $g$ is the spectroscopic
splitting factor, and $\mu_{B}$ is the Bohr magneton. Here
$\mathbf{A=A(r)}$ is the coupling between the muon and electron
spin, which we assume depends on the distance between them,
although at present the origin of this coupling is unknown. In a
mean field approximation, $\mathbf{S\rightarrow}$ $\left\langle
\mathbf{S}\right\rangle =\chi\mathbf{H}/g\mu_{B}$. Therefore, the
muon experiences a magnetic field
$\mathbf{B}=(1+\mathbf{A\chi})\mathbf{H}$. Assuming that the
susceptibility and the couplings are isotropic, the time evolution
of a muon spin due to static field inhomogeneity is given by:
\begin{equation}
P_{TF}(t)=\int\cos[\gamma_{\mu}(1+\{\left\langle A\right\rangle +\delta
A\}\chi)Ht]\rho(\delta A)d\delta A \label{Rlx_Theory}%
\end{equation}
where $\left\langle A\right\rangle $ is the mean coupling,
$\gamma_\mu=85.162~MRad/(T\cdot sec)$ is the muon gyromagnetic
ratio and $\rho$ is the distribution of coupling variations. From
Eqs.~\ref{TFFit} and \ref{Rlx_Theory} we find, using inverse
Fourier transform that: (I) The frequency shift is given by
\begin{equation}
K=\frac{\omega_0}{\gamma_\mu H}\left\langle A\right\rangle \chi. \label{KtoChi}%
\end{equation}
(II) The distribution $\rho$ can be expressed as
\begin{equation}
\rho(\delta A)=f(\sigma/\{2\pi|\delta A|\})/(2|\delta A|) \label{rho}%
\end{equation}
where the effective width of the distribution, $\sigma$, is given by
\begin{equation}
\sigma=1/\left\vert T_{2}\chi\gamma_{\mu}H\right\vert , \label{Sigma}%
\end{equation}

\begin{equation}
f(x)=\sqrt{x}\left[  \sin(\pi x/2)\left\{  1-2FS(\sqrt{x})\right\}  +\cos(\pi
x/2)\left\{  1-2FC(\sqrt{x})\right\}  \right]  \label{fx}%
\end{equation}
and $FS$ and $FC$ are the Fresnel functions \cite{Thesis}. Thus,
according to Eq.~\ref{Sigma}, if the ratio between $T_{2}^{-1}$
and $K$ remains constant as the temperature decreases, the most
likely conclusion is that $\sigma$ is temperature independent, and
no lattice deformations are present. This conclusion would be
wrong only if by accident $\left\langle A\right\rangle /(T_{2}K)$
turns out to be temperature independent while $(T_{2}K)^{-1}$ does
vary with $T$. To eliminate this possibility $\chi$ must be
measured directly as discussed below. It should be pointed out
that the same calculation could be done by assuming that $\chi$ is
distributed instead of $A$. In this case a proportionality between
$T_2^{-1}$ and $K$ would mean a temperature independent
distribution of $\chi$ throughout the lattice. This would equally
role out lattice distortion upon cooling.

\begin{figure}[ptb]
\begin{center}
\includegraphics{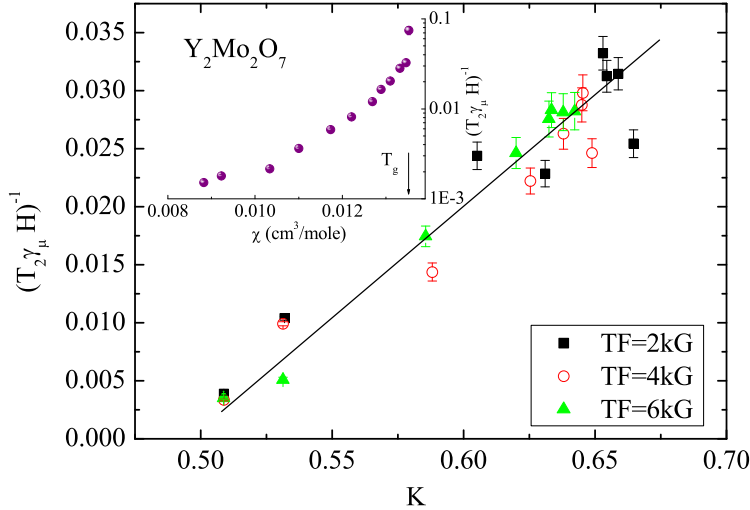}
\end{center}
\caption{Demonstrating the linear relation between the muon spin
relaxation rate normalized by the field
$(T_{2}\gamma_{\mu}H)^{-1}$ versus the shift in muon spin rotation
frequency $K$, which is proportional to the susceptibility $\chi$.
The
inset shows the same type of measurement on a semi log scale in the pyrochlore compound Y$_{2}%
$Mo$_{2}$O$_{7}$ taken from Ref.~\cite{eva}. } \label{T2andK}
\end{figure}

In Fig.~\ref{T2andK} we show $(T_{2}\gamma_{\mu} H)^{-1}$ versus
$K$ with the temperature as an implicit parameter. The two
quantities are linearly dependent at all fields. Allowing for a
base line shift, which does not originate in localized spins will
lead to a
proportionality relation between these quantities.  This stands in strong contrast to Y$_{2}%
$Mo$_{2}$O$_{7}$ where the muon transverse relaxation grows as a
function of $\chi$ faster than exponentially. This is demonstrated
in the inset of Fig.~\ref{T2andK} using data from Ref. \cite{eva}
on a semi log scale. We conclude that in Tb$_{2}$Ti$_{2}$O$_{7}$
the muon transverse
relaxation has the same temperature dependence as the shift. %
In fact, by
calculating $\sigma$ for each data point using Eq. \ref{Sigma} we find that
$\Delta\sigma/\overline{\sigma}$ is 15\% for Tb$_{2}$Ti$_{2}$O$_{7}$ and 115\%
for Y$_{2}$Mo$_{2}$O$_{7}$ where $\overline{\sigma}$ and $\Delta\sigma$ are
the average and standard deviation of $\sigma$ respectively. $\Delta
\sigma/\overline{\sigma}$ is a measure of the relative change in the distances
variations due to temperature changes.%

We also performed DC-susceptibility measurements using a cryogenic
SQUID magnetometer, and home built Faraday balance
\cite{sakakibara} in a Dilution Refrigerator (DR). In the Faraday
balance the powder sample is attached to a metallic sample holder,
which serves as one of the plates of a capacitor, and is suspended
on a spring. The sample experience both a field $H$ and a field
gradient of $1~T/m$. The capacitance $C$ is measured as a function
of $H$ and $\chi\propto dC^{-1}/dH$.

\begin{figure}
[ptb]
\begin{center}
\includegraphics{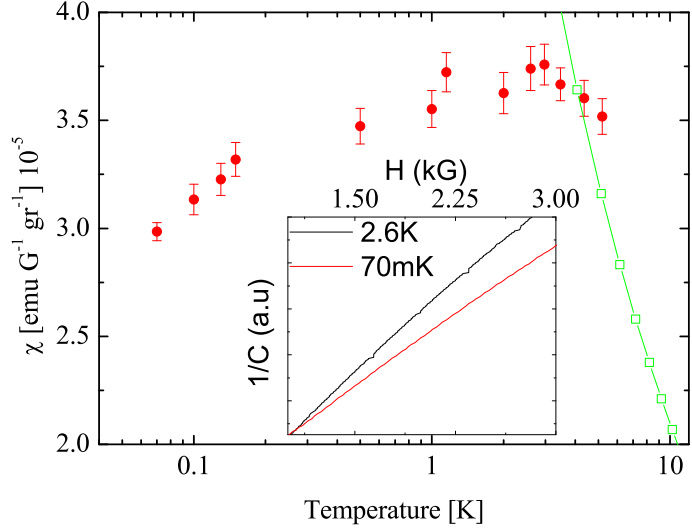}%
\caption{The DC-susceptibility as measured by SQUID, and Faraday balance
mounted in a dilution refrigerator. The two measurements do not overlap
probably due to the difficulty of cooling the large amount of powder in the
balance. Nevertheless a decrease in $\chi$ is observed as the temperature
decreases. Inset shows raw inverse capacitance versus applied field. The slope
is proportional to $\chi$.}%
\label{susc}%
\end{center}
\end{figure}

In the inset of Fig.~\ref{susc} we show raw data of $C^{-1}$ versus $H$ at
70mK and 2.6K. At all fields the slope at 70~mK is smaller than at 2.6~K. This
means that at some temperature, in between these two, the susceptibility
starts to decrease. In Fig. \ref{susc} we show the temperature dependent
susceptibility by both experimental methods. The two measurements do not
overlap in the 2-5 K region. We attribute this problem to difficulties in
cooling of the entire volume of the sample in the DR. In other words, the mean
temperature of the sample $T_{s}$ is higher than the DR temperature $T_{DR}$.
Nevertheless, as $T_{DR}$ decreases, and so does $T_{s}$, the susceptibility
does not increase. In fact, below $T_{DR}\sim0.2K$, the susceptibility
decreases. A saturated susceptibility and the decrease below $\sim0.2$~K is in
rough agreement with previous measurements of Luo \textit{et al. }\cite{Luo}.
This result shows qualitatively that $T_{2}^{-1}$ and $\chi$ do not vary
substantially with $T$ below $2$~K, which supports our previous conclusions.

To summarize, we compare the transverse relaxation rate resulting from
internal field distribution to susceptibility measured by the shift in the
muon rotation frequency and DC susceptibility. We find that the relaxation
rate has the same temperature dependence as the susceptibility. This indicates
that the only reason for increasing relaxation upon cooling is an increase in
the electronic moment size. Therefore, there is no evidence for lattice deformation in Tb$_{2}%
$Ti$_{2}$O$_{7}$ that is static on the time scale of 0.1 $\mu$sec.

\ack
 We are grateful to the machine and instrument groups at Paul
Scherrer Institute, Switzerland, whose outstanding efforts have
made these experiments possible. The authors wish to acknowledge
the financial support of NATO through a collaborative linkage
grant and the support from the European Science Foundation (ESF)
'Highly Frustrated Magnetism'.

\end{document}